\documentclass[aps,prl,twocolumn,amsmath,amssymb,superscriptaddress,nofootinbib,showpacs,preprintnumbers]{revtex4-1}
\usepackage{aas_macros}
\usepackage{graphicx}
\usepackage{epstopdf}
\usepackage[colorlinks=true]{hyperref}
\usepackage[usenames]{color}
\usepackage[normalem]{ulem} 

\newcommand{\red}{\textcolor{red}}
\newcommand{\blue}{\textcolor{blue}}

\def\fit#1{\red{#1}}
\def\given#1{\blue{#1}}

\begin{document}

\title{New test of Lorentz invariance using the MICROSCOPE space mission}

\author{H\'el\`ene Pihan-le Bars} 
\affiliation{SYRTE, Observatoire de Paris, Universit\'e PSL, CNRS, Sorbonne Universit\'e, LNE, 75014 Paris, France}

\author{Christine Guerlin}
\affiliation{SYRTE, Observatoire de Paris, Universit\'e PSL, CNRS, Sorbonne Universit\'e, LNE, 75014 Paris, France}
\affiliation{Laboratoire Kastler Brossel, ENS-Universit\'e PSL, CNRS, Sorbonne Universit\'e, Coll\`ege de France, 75005 Paris, France}

\author{Aur\'elien Hees}
\affiliation{SYRTE, Observatoire de Paris, Universit\'e PSL, CNRS, Sorbonne Universit\'e, LNE, 75014 Paris, France}

\author{Romain Peaucelle}
\affiliation{SYRTE, Observatoire de Paris, Universit\'e PSL, CNRS, Sorbonne Universit\'e, LNE, 75014 Paris, France}
\affiliation{Ecole Sup\'erieure des Techniques A\'eronautiques et de Construction Automobile (ESTACA), 78066 Saint-Quentin-en-Yvelines, France}

\author{Jay D. Tasson}
\affiliation{Department of Physics and Astronomy, Carleton College, Northfield, MN 55057, USA}

\author{Quentin G. Bailey}
\affiliation{Department of Physics and Astronomy, Embry-Riddle Aeronautical University, Prescott, AZ 86301, USA}

\author{Geoffrey Mo}
\affiliation{Department of Physics and Astronomy, Carleton College, Northfield, MN 55057, USA}

\author{Pac\^ome Delva}
\affiliation{SYRTE, Observatoire de Paris, Universit\'e PSL, CNRS, Sorbonne Universit\'e, LNE, 75014 Paris, France}

\author{Fr\'ed\'eric Meynadier}
\affiliation{SYRTE, Observatoire de Paris, Universit\'e PSL, CNRS, Sorbonne Universit\'e, LNE, 75014 Paris, France}
\affiliation{Bureau International des Poids et Mesures, Pavillon de Breteuil, 92312 S\`{e}vres, France}

\author{Pierre Touboul}
\affiliation{DPHY, ONERA, Universit\'e Paris Saclay, 92322 Ch\^atillon, France}

\author{Gilles M\'etris}
\affiliation{Universit\'e C\^ote d’Azur, Observatoire de la C\^ote d'Azur, CNRS, IRD, G\'eoazur, 06560 Valbonne, France}

\author{Manuel Rodrigues}
\affiliation{DPHY, ONERA, Universit\'e Paris Saclay, 92322 Ch\^atillon, France}

\author{Jo\"el Berg\'e}
\affiliation{DPHY, ONERA, Universit\'e Paris Saclay, 92322 Ch\^atillon, France}

\author{Peter Wolf}\email{peter.wolf@obspm.fr}
\affiliation{SYRTE, Observatoire de Paris, Universit\'e PSL, CNRS, Sorbonne Universit\'e, LNE, 75014 Paris, France}

\date{\today}


\begin{abstract}
We use data from the T-SAGE instrument on board the MICROSCOPE space mission to search for Lorentz violation in matter-gravity couplings as described by the Lorentz violating Standard-Model Extension (SME) coefficients $(\bar{a}_\text{eff})_\mu^w$, where ($\mu = T,X,Y,Z$) and ($w = e,p,n$) for the electron, proton and neutron. One of the phenomenological consequences of a non-zero value of those coefficients is that test bodies of different composition fall differently in an external gravitational field. This is similar to ``standard'' tests of the universality of free fall, but with a specific signature that depends on the orbital velocity and rotation of the Earth. We analyze data from five measurement sessions of MICROSCOPE spread over a year finding no evidence for such a signature, but setting constraints on linear combinations of the SME coefficients that improve on best previous results by one to two orders of magnitude. Additionally, our independent linear combinations are different from previous ones, which increases the diversity of available constraints, paving the way towards a full decorrelation of the individual coefficients.


\end{abstract}

\maketitle

\section{Introduction}

\textit{Introduction --} The Einstein Equivalence Principle is the foundation of General Relativity and all metric theories of gravitation. It includes - among others - the Weak Equivalence Principle (WEP) and Local Lorentz Invariance (LLI)\cite{will1993}. The WEP, which states that the universality of free fall for bodies with negligible self-gravity, has been tested recently by the MICROSCOPE space mission with a sensitivity of $< 2 \times 10^{-14}$ to the E\"otv\"os parameter \cite{Touboul2017}. MICROSCOPE compared the differential acceleration of two test-masses of different composition (Ti and Pt). No composition-dependent deviation from geodesic motion was found, but this first analysis has improved by about one order of magnitude the previous constraints obtained from torsion balance experiments and  Lunar Laser Ranging (LLR) \cite{schlamminger2008,Wagner2012,Williams2012,Viswanathan2018}.

The MICROSCOPE mission also offers a valuable opportunity to constrain the matter-gravity sector of the Standard-Model Extension (SME), an effective field theory developed to characterize low-energy signatures of Planck-scale physics \cite{
kostelecky1995,colladay1997,kostelecky1989,colladay1998,kostelecky2011matter}, and particularly Lorentz invariance violations \cite{kostelecky1989}. The SME is a general framework allowing for systematic searches for LLI violation. The latter is quantified by SME tensor fields (more precisely their vacuum expectation values, called coefficients) that parameterize the amplitude of the LLI violation. 

Different combinations of coefficients can be probed effectively
by different physical systems.
Hence the SME has been used to explore
LLI across a large range of
phenomena.
While we refer the reader to Ref.\ \cite{kostelecky-2008}
for an annually-updated review of experimental and observational progress
and a full list of references,
we briefly summarize the breadth of these efforts.
Nongravitational SME tests
have been performed with a wide range of systems
including
atomic clocks \cite{Wolf2006,Pihan-LeBars2017,Hohensee2013,Flambaum2017}, 
comagnetometers \cite{Smiciklas2011},
neutrino \cite{aharmim2018} and meson \cite{aaij2016} oscillations,
quark production \cite{abazov2012},
muon $g-2$ experiments\cite{bennett2008},
torsion pendula \cite{heckel2006},
particle traps \cite{smorra2017},
resonant cavities \cite{baynes2012},
and astrophysical photon propagation \cite{friedman2019}.
Complementary progress has been made via many gravitational SME searches \cite{[{For a review, see, }]Hees2016}
including tests with 
gravimeters \cite{flowers2017,cgshao2018},
spin precession \cite{tasson2012},
solar-system data \cite{kostelecky2011matter,Hees2015},
LLR \cite{Bourgoin2016,Bourgoin2017},
gravitational waves \cite{gwgrb2017},
WEP experiments \cite{kostelecky2011matter,Hohensee2013b,hohensee2011},
pulsar timing \cite{lshao2019_mat,lshao2019},
short-range gravity \cite{cgshao2016},
and very long baseline interferometry \cite{LePoncin-Lafitte2016}.

In this work we focus on matter-gravity couplings \cite{kostelecky2011matter}, \textit{i.e.} to what extent the behavior of a test body in a gravitational field is affected by couplings to Lorentz-violating background fields. 
The WEP tests are novel
among gravitational tests for their ability to distinguish 
the species-dependent coefficients associated with matter-gravity couplings
from the universal gravity-sector coefficients.
Moreover,
relative measurements on colocated test particles offer higher-precision
tests compared with other observables in which Lorentz violation in matter-gravity couplings has been sought \cite{kostelecky2011matter,flowers2017,tasson2012,Bourgoin2017,lshao2019_mat,Hees2015}.
This combination makes MICROSCOPE an ideal system in which to search for these effects.
We are sensitive to violations arising from the composition dependent $(\bar{a}_\text{eff})_\mu$ coefficient, which has been the primary target
of searches in matter-gravity couplings to date.
Our sensitivity is independent of the universal gravitational coefficient $\bar{s}_{\mu\nu}$,\footnote{$({a}^B_{\text{eff}})_\mu$ and ${s}_{\mu\nu}$ are SME fields for which SME coefficients $(\bar{a}^B_{\text{eff}})_\mu$ and $\bar{s}_{\mu\nu}$ provide vacuum expectation values in spontaneous Lorentz violation scenarios.}
with which it has been fully correlated in many of the other high-sensitivity
searches such as LLR tests \cite{Bourgoin2017}.

We present the first results of a LLI test with the MICROSCOPE space mission, searching for a putative non-zero value of $(\bar{a}_\text{eff}^w)_\mu$ for the fundamental atomic particles ($w=e,p,n$). We analyzed five measurement sessions spread over 2017 to search for an orientation dependent differential acceleration of the Pt vs. Ti test masses. As the $(\bar{a}_\text{eff})_\mu$ vector is constant in a sun centered non-rotating frame \cite{kostelecky2011matter} , the dependence of the expected acceleration on the position and orientation of the instrument is more complex than for the ``simple" WEP test, hence a specific data analysis was necessary, that is complementary to ref. \cite{Touboul2017}. 

\textit{Theoretical model --} We developed a theoretical model that allows us to extract the values of $(\bar{a}_\text{eff})_\mu$ from the differential acceleration measurements and in-flight data (orbit, attitude, gravity and gravity gradients, temperatures, ...). 
The MICROSCOPE satellite was in a heliosynchronous circular orbit at 710 km altitude and spinning around an axis (the y-axis in the instrument frame) that is perpendicular to the orbital plane. We use measurements of the differential acceleration of the two test masses along the x-axis, which is the most sensitive axis of the instrument. For all details on the mission and the T-SAGE instrument see \cite{Touboul2017} and references therein.

The contributions to the SME action for a body B of mass $m^B$ that are relevant here take the form (see \cite{kostelecky2011matter} for details)
\begin{equation}\label{equ:action}
S^B \simeq \int d\lambda ( -m^Bc\sqrt{-g_{\mu\nu} u^\mu u^\nu} - ({a}^B_{\text{eff}})_\mu u^\mu/c) \, ,
\end{equation}
where $c$ is the speed of light, $g_{\mu\nu}$ is the metric tensor, $u^\mu \equiv \mathrm{d}x^\mu/\mathrm{d}\lambda$ is the four velocity of B, $\lambda$ parameterized the path of B, and 
$({a}^B_{\text{eff}})_\mu$
is a composition-dependent field that vanishes when LLI is satisfied. 
The effective coefficient $(\bar{a}_{\text{eff}})_\mu$ is a combination of $\bar{a}_\mu$ and $\bar{e}_\mu$ coefficients $(\bar{a}_{\text{eff}})_\mu \equiv \bar{a}_\mu - mc^2 \bar{e}_\mu$, and is generally written with a numerical factor $\alpha$ that depends on the specifics of the theory \cite{kostelecky2011matter}. Conventionally constraints are given directly for $\alpha(\bar{a}_{\text{eff}})_\mu$.

We use the Lagrangian for a test mass in the field sourced by the Earth that is obtained from (\ref{equ:action}) (see  \cite{kostelecky2011matter} Sec. VII.A. for details) to derive a model that relates the MICROSCOPE differential acceleration measurements and in-flight data to $\alpha(\bar{a}^{(d)}_{\text{eff}})_\mu \equiv \alpha(\bar{a}^{B1}_{\text{eff}})_\mu/m^{B1} - \alpha(\bar{a}^{B2}_{\text{eff}})_\mu/m^{B2}$,
\onecolumngrid
\begin{center}
\begin{align}
\label{eq_mes_mic_sme}
\given{\gamma_{\hat{x}}} =& \,\fit{b} + \given{\text{ S}_{\hat{\text{x}}\hat{\text{x}}}} \fit{\Delta_{\hat{\text{x}}}} + \given{( \text{S}_{\hat{\text{x}}\hat{\text{y}}}}+ \given{\dot{\Omega}_z ) \Delta_{\hat{\text{y}}}}+ (\given{\text{S}_{\hat{\text{x}}\hat{\text{z}}}}-\given{\dot{\Omega}_y} ) \fit{\Delta_{\hat{\text{z}}}}+ 2\given {g_{\hat{\text{x}}}}\big[\fit{\alpha ( \bar{a}^{\text{(d)}}_{\text{eff}})_{T}} +\given{\beta_{\text{X}}} \fit{\alpha ( \bar{a}^{\text{(d)}}_{\text{eff}})_{X}}+\given{\beta_{\text{Y}}}  \fit{\alpha ( \bar{a}^{\text{(d)}}_{\text{eff}})_{Y}} +\given{\beta_{\text{Z}}} \fit{\alpha ( \bar{a}^{\text{(d)}}_{\text{eff}})_{Z}} \big] \nonumber \\
-& \dfrac{6 G M_{\oplus} R_{\oplus}^2 }{5c\given{r}^{5}} \left(\given{R_{\hat{x}\tilde{x}}\tilde{x}^{\text{orb}}} + \given{R_{\hat{x}\tilde{y}}\tilde{y}^{\text{orb}}} + \given{R_{\hat{x}\tilde{z}}\tilde{z}^{\text{orb}}}\right) \big[\given{\tilde{x}^{\text{orb}}}\fit{\alpha ( \bar{a}^{\text{(d)}}_{\text{eff}})_{Y}} - \given{\tilde{y}^{\text{orb}}}\fit{\alpha( \bar{a}^{\text{(d)}}_{\text{eff}})_{X}} \big] \omega_{\tilde{\text{z}}} \nonumber\\
+& \dfrac{2 G M_{\oplus} R_{\oplus}^2}{5c\given{r}^{3}}\big[\fit{\alpha ( \bar{a}^{\text{(d)}}_{\text{eff}})_{Y}} \given{R_{\hat{x}\tilde{x}}}-\fit{\alpha ( \bar{a}^{\text{(d)}}_{\text{eff}})_{X}} \given{R_{\hat{x}\tilde{y}}} \big] \omega_{\tilde{\text{z}}} \,,
\end{align}
\end{center}
\twocolumngrid
\noindent where quantities in red need to be estimated in our data analysis by fitting the model to the data, and quantities in blue are obtained from the mission center and INPOP planetary ephemerides \cite{viswanathan2017inpop17a}.

Three coordinate systems are used in this model: the instrument frame $\hat{x}_{\mu}$, the geocentric frame (GCRF) $\tilde{x}_{\mu}$, and Sun centered frame $X_{\mu}$. The latter two coordinate systems are kinematically non-rotating \cite{Soffel2003}. The data provided by the mission data center are: the differential acceleration $\gamma_{\hat{x}}$, the satellite attitude given by $R_{\hat{\mu}\tilde{\mu}}$ the rotation matrix from the GCRF to the instrument frame, the orbital position of the satellite $\tilde{x}^{\text{orb}}$, the gradient of the Earth's gravitational potential in the instrument frame $g_{\hat{x}}$ and the satellite angular acceleration $\dot{\Omega}_{\hat{x}}$. We also use the gravity gradient tensor $ \text{T}_{\hat{\text{x}}\hat{\text{x}}}$, and the satellite angular velocity matrix $ \Omega_{\hat{\text{x}}\hat{\text{x}}}$ \cite{Touboul2017}, introduced in the model through the matrix $\text{S}_{\hat{\text{x}}\hat{\text{x}}} = \text{T}_{\hat{\text{x}}\hat{\text{x}}}+ \Omega_{\hat{\text{x}}\hat{\text{x}}}^2$. The first four terms of the model depend on the off-centering of the test masses along $\hat{x}, \hat{y}$ and $\hat{z}$ axes - respectively $\Delta_{\hat{\text{x}}}, \Delta_{\hat{\text{y}}}$ and $\Delta_{\hat{\text{z}}}$ - and on an overall bias $b$, all of unknown amplitude \cite{Touboul2017}. The parameters $\Delta_{\hat{\text{x}}}, \Delta_{\hat{\text{y}}}, b$ need to be estimated together with the SME coefficients in order to correctly take into account any correlations between the SME parameters and these ``technical" ones. The offcentering $\Delta_{\hat{\text{y}}}$ is less critical as the satellite is spinning around the $\hat{\text{y}}$ axis, which is perpendicular to the orbital plane, thus the coefficient of the $\Delta_{\hat{\text{y}}}$ term is much smaller. It is obtained from dedicated calibration sessions and provided by the mission centre. Finally, $G M_{\oplus},R_{\oplus},\omega_{\tilde{i}}$ are the gravitational parameter, mean radius, and angular velocity of the Earth, $r^2 = (\tilde{x}^{\text{orb}})^2+(\tilde{y}^{\text{orb}})^2+(\tilde{z}^{\text{orb}})^2$, and $\beta_{\rm I} \equiv v_{\rm I}/c$ is the Earth's orbital velocity obtained from INPOP planetary ephemerides. The basic sampling interval of all data files is 0.25~s except the orbit data (1~min) and temperature data (1~s), which we interpolated to 0.25~s.

Note that (\ref{eq_mes_mic_sme}) is expressed in terms of a differential SME coefficient, which following the method described in Ref.\ \cite{kostelecky2011matter} Sec.\ VI.B. and taking into account the isotopic composition of the Pt:Rh and Ti:Al:V alloys used in MICROSCOPE \cite{Touboul2017}, is
$\alpha (\bar{a}_{\text{eff}}^{\text{(d)}})_\mu = A\alpha(\bar{a}_{\text{eff}}^{\text{(n-e-p)}})_\mu,$
where $(\bar{a}_{\text{eff}}^{\text{(n-e-p)}})_\mu \equiv (\bar{a}_{\text{eff}}^{\text{n}})_\mu - (\bar{a}_{\text{eff}}^{\text{e}})_\mu- (\bar{a}_{\text{eff}}^{\text{p}})_\mu$, and $A \simeq 0.06$~GeV$^{-1}$.

%
The modulations of the Lorentz violation signal are mainly due to the oscillation of $g_{\hat{x}}$ at $f_{\text{EP}} = f_{\text{orb}} +  f_{\text{spin}}$ ($f_{\text{EP}} \approx 3.1$~mHz, $f_{\text{orb}} \approx 0.17$~mHz  \cite{Touboul2017}), where $f_{\text{orb}}$ and $f_{\text{spin}}$ are respectively the orbital and rotational frequency of the satellite. Additionally, the model also has an annual modulation mainly via the dependence on the Earth's orbital velocity ($\beta_X$, $\beta_Y$, $\beta_Z$). Finally, additional modulations at $f_{\text{orb}}$ arise from the small terms in the second line of (\ref{eq_mes_mic_sme}). In summary, the frequencies involved in the SME model are : $\left\lbrace f_\text{spin}, f_\text{spin} + f_\text{orb}, f_\text{spin} + 2f_\text{orb}\right\rbrace$ each of which are affected by annual sidebands.

\begin{figure}[h]
\centering
\includegraphics[width=0.5\textwidth]{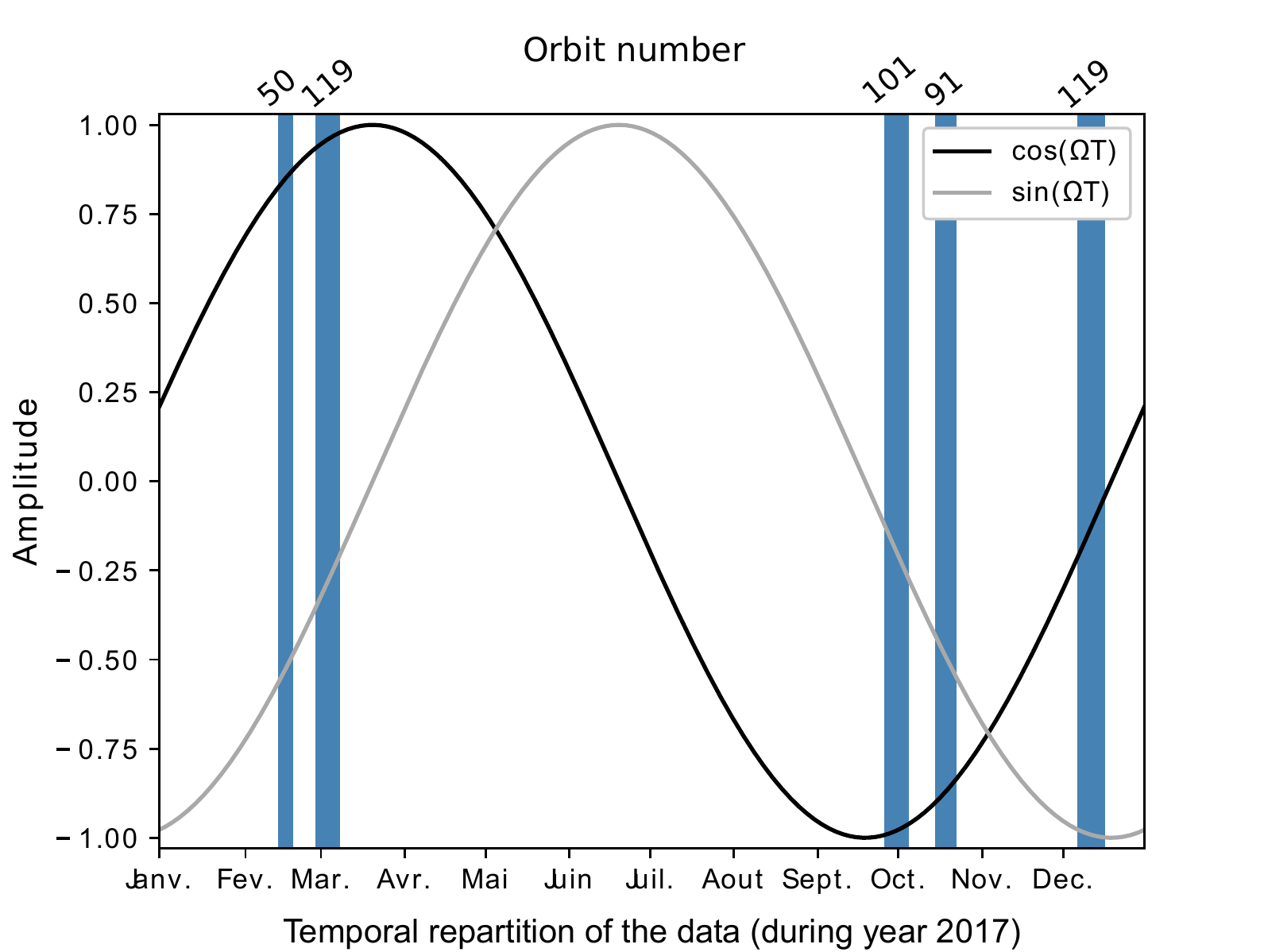}
\caption{Temporal distribution over the year of the measurements sessions used for our analysis. Numbers on top indicate the number of complete satellite orbits in the session. The sine curves indicate the form of the temporal evolution of $\beta_X$ ($\propto$~sin$(\Omega T)$) and $\beta_Y, \beta_Z$ ($\propto$~cos$(\Omega T)$) used in (\ref{eq_mes_mic_sme}), where $\Omega$ is the annual frequency.}
\label{fig:data}
\end{figure}

\textit{Data and statistical analysis --} Our data consists of 5 measurement sessions (nos. 210, 218, 326, 358, 404) with durations ranging from 4 to 8 days, spread over Feb. to Sept. 2017 (see fig. \ref{fig:data}). Whilst this obviously implies large dead times between sessions, the proportion of missing or corrupted data within each session was remarkably low ($< 10^{-5}$). One of the sessions (no. 218, Feb. 2017) was the same one as used in \cite{Touboul2017}, which allowed cross-checks (see below).

\begin{figure}[h]
\includegraphics[width=0.43\textwidth]{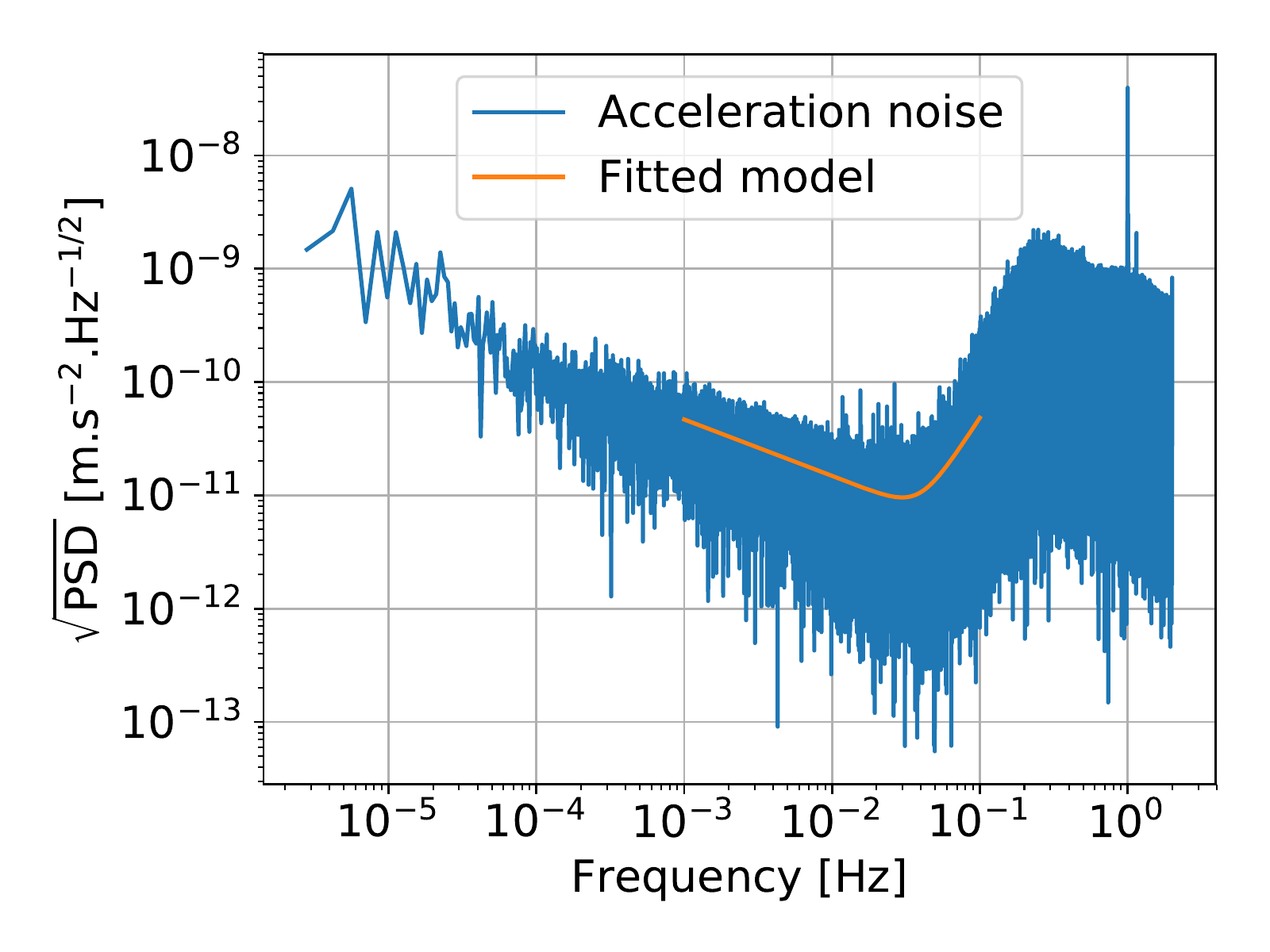}
\caption{$\sqrt{\rm{PSD}}$ of the acceleration measurement residuals in blue, and the two-slopes model (see text) in orange with amplitudes $a_{-1} = 2.2\times 10^{-24}$~(m~s$^{-2}$)$^2$ and $a_{4} = 2.3\times 10^{-17}$~(m~s$^{-2}$)$^2$Hz$^{-5}$.}
\label{fig:acc}
\end{figure}

Considering the large dead times between sessions and the complexity of the noise (see fig. \ref{fig:acc}), we worked in the time domain using a Least-Squares Monte-Carlo (LSMC) analysis. This method is an efficient and simple way to deal with coloured noise and gaps \cite{Delva2018,Savalle2019} by simulating a large number ($N \sim 10000$) of synthetic data sets with the same noise characteristics and gaps as the real one and least squares fitting the model to each of them. The values of the fit parameters are obtained from the real data set, their statistical uncertainties and correlations from the $N$ simulated ones.   

The raw differential acceleration ($\gamma_{\hat{x}}$) data is first high pass filtered by removing a polynomial of order 5 to account for slow drifts and variations. We do this in a first step for simplicity, but we have verified that fitting the polynomial together with our model (\ref{eq_mes_mic_sme}) does not change our results, nor does using a higher order polynomial. The noise can be modelled by a sum of a low frequency pink noise, with a $f^{-1}$ slope, and a high frequency noise with a $f^4$ slope. The first one is the thermal noise of the gold wire connecting the test masses to the outer cage \cite{Touboul2017}, the second one is the second derivative of the position measurement (white) noise. The corresponding power spectral density (PSD) model $S_\gamma(f) = a_{-1} f^{-1} + a_4 f^4$ is fitted to the PSD of the data residuals after the fit of (\ref{eq_mes_mic_sme}). The obtained values of $a_{-1}$ and $a_4$ for each session are then used to generate the synthetic data sets. Figure \ref{fig:acc} shows the $\sqrt{{\rm PSD}}$ of the residuals from one session (no. 404, July 2017) together with the best fit model. The agreement is satisfactory in the region of interest around $3.1 \pm 0.2$~mHz where our signal is located.

\begin{figure}[h]
\includegraphics[width=0.43\textwidth]{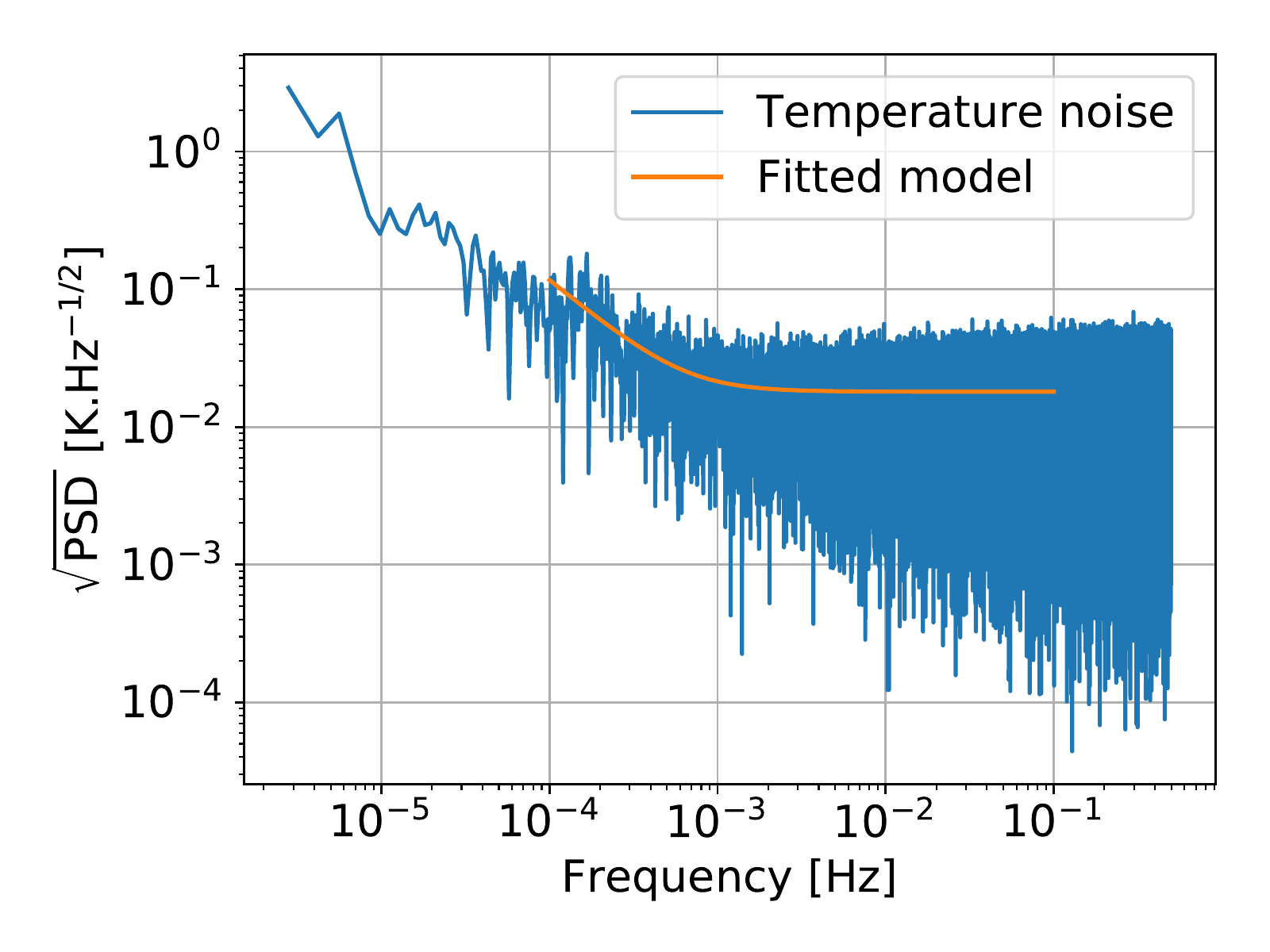}
\caption{$\sqrt{\rm{PSD}}$ of the temperature measurement residuals in blue, and the two-slopes model in orange (see text) with amplitudes $d_{-2} = 1.3\times 10^{-10}$~K$^2$Hz and $d_{0} = 3.3\times 10^{-4}$~K$^2$Hz$^{-1}$.}
\label{fig:temp}
\end{figure}

\textit{Systematic effects --} As discussed in detail in \cite{Touboul2017} the by far dominant systematic effect in the frequency region of interest is related to thermal fluctuations that give rise to corresponding fluctuations of $\gamma_{\hat{x}}$. The corresponding coupling coefficient between the temperature of the instrument and the differential acceleration was determined by dedicated measurement sessions where the baseplate temperature was varied intentionally \cite{Touboul2017}, giving $\gamma_{\hat{x}}=C \Delta T$ where $\Delta T$ is the baseplate temperature fluctuation and $C = 4.3\times 10^{-9}$~m~s$^{-2}$K$^{-1}$. To determine the corresponding effect on our parameter estimations we used the baseplate temperature data, ``converted" it to acceleration data using $C$, and analysed it using the same LSMC method as for the acceleration data. The resulting parameters and their uncertainties (to be conservative we used the quadratic sum of the two) are then our estimate of the systematic uncertainties. The noise model used for the temperature data $S_{\Delta T}(f) = d_{-2} f^{-2} + d_0$ contains white measurement noise of the thermistors and a $f^{-2}$ component that could be a random walk temperature noise. Figure \ref{fig:temp} presents temperature data residuals (after removal of slow drifts by a 3rd order polynomial) from one session (July 2017) and the best fit model, showing good agreement in the region of interest.

\textit{Results --} We first checked our analysis method by comparing it to previous work \cite{Touboul2017}. To do so, we analysed the same single session as \cite{Touboul2017} (no. 218) and simplified our model (\ref{eq_mes_mic_sme}) by setting all SME parameters to zero except $\alpha (\bar{a}_{\text{eff}}^{\text{(d)}})_\text{T}$. Then the model is identical to a standard WEP test, with the simple correspondence $\delta = 2 \alpha (\bar{a}_{\text{eff}}^{\text{(d)}})_\text{T}$, where $\delta$ is the the E\"otv\"os parameter. The results are summarized in table \ref{table_ep}. 
\begin{table}[h]
\centering
$\begin{array}{|c|c|c|c|}
\hline
 \text{Parameter} &\text{Value and uncertainties}&\text{Units}\\
 \hline
 \delta      &     (4.0 \pm 9.6_{\rm stat} \pm 13.0_{\rm syst}) \times 10^{-15}& -\\
 \Delta_x &   ( 20.2 \pm 0.04) \times 10^{-6}   & m \\
 \Delta_z  &  ( -5.77 \pm 0.04)\times 10^{-6}     & m \\
\hline
\end{array}$
\caption{Estimations of the E\"otv\"os parameter $\delta$ and the offcenterings for session 218. For the offcenterings only statistical uncertainties are shown. All correlation coefficients are $\leq 0.08$.}
\label{table_ep}
\end{table}
Our results agree, within their uncertainties, with the ones of \cite{Touboul2017}, which are $\delta = (-1 \pm 9_{\rm stat} \pm 9_{\rm syst}) \times 10^{-15}$. Our slightly larger uncertainties could be due to the LSMC estimator being non-optimal and/or slight discrepancies of our noise models or evaluation of the systematics. Nonetheless, the good agreement between two independent analyses is a valuable and conclusive cross-check.

For the global SME analysis we weight the data of each session by the inverse of the square of the total uncertainty on $\delta$ obtained for each individual session, which are $(3.8,1.6,2.7,1.6,1.0)\times 10^{-14}$ for sessions no. $(210, 218, 326, 358, 404)$. We carry out a global LSMC fit of (\ref{eq_mes_mic_sme}) using those weights, obtaining estimates, uncertainties, and correlations for the four 
combinations $\alpha (\bar{a}_{\text{eff}}^{\text{(d)}})_\mu$, 10 offcenterings (2 per session), and 5 biases. The results for the SME coefficients are given in Tab.\ref{res_comb}.
\begin{table}[h]
\centering
$\begin{array}{|c|c|}
\hline
\text{Coefficient}& \text{Value and uncertainties [GeV]}\\
 \hline
\alpha(\bar{a}_{\text{eff}}^{\text{(n-e-p)}})_T      & \left(6.3 \pm  12\right) (10)(6.0) \times 10^{-14} \\
\alpha(\bar{a}_{\text{eff}}^{\text{(n-e-p)}})_X     & \left(0.81 \pm 1.7 \right)(1.4)(0.98) \times 10^{-9}  \\
\alpha(\bar{a}_{\text{eff}}^{\text{(n-e-p)}})_Y      &\left(0.67\pm 3.1 \right)  (1.4)(2.7) \times 10^{-7}  \\
\alpha(\bar{a}_{\text{eff}}^{\text{(n-e-p)}})_Z     & \left(-1.55 \pm  7.1 \right)(3.2)(6.3) \times 10^{-7} \\
\hline
\end{array}$
\caption{SME coefficients obtained from the global analysis of five sessions. 2nd and 3rd brackets show statistical and systematic uncertainties (68\% confidence) respectively. Correlation coefficients are $\sim$~0.9 between T and X components, $\sim$~1 between Y and Z, and $\leq$~0.02 otherwise.}
\label{res_comb}
\end{table}  

Given the large correlations between the SME coefficients we also perform a singular value decomposition (SVD) of the covariance matrix to determine uncorrelated linear combinations of coefficients (see appendix C of \cite{Pihan-LeBars2017} for details). The resulting linear combinations $a_1, a_2, a_3, a_4$ are given in Tab.\ref{tab:lin-comb}, and their values and uncertainties in Tab.\ref{tab:lin-comb-res}.
\begin{table}[h]
\centering
$\begin{array}{ccccc}
\hline\hline
& \alpha(\bar{a}_{\text{eff}}^{\text{(n-e-p)}})_T & \alpha(\bar{a}_{\text{eff}}^{\text{(n-e-p)}})_X & \alpha(\bar{a}_{\text{eff}}^{\text{(n-e-p)}})_Y & \alpha(\bar{a}_{\text{eff}}^{\text{(n-e-p)}})_Z  \\
\hline
a_1 & 1.0 & -6.0\,\,10^{-5} & 4.8\,\,10^{-6} & 2.0\,\,10^{-6} \\
a_2 & 5.9\,\,10^{-5} & 0.99 & 0.11 & 0.050 \\
a_3 & -1.3\,\,10^{-5} & -0.12 & 0.91 & 0.39 \\
a_4 & 1.2\,\,10^{-9} & -4.9\,\,10^{-5} & -0.40 & 0.92 \\
\hline\hline
\end{array}$
\caption{Composition of the independent linear combinations of $\alpha(\bar{a}_{\text{eff}}^{\text{(n-e-p)}})_\mu$ coefficients obtained using a SVD of their covariance matrix.}
\label{tab:lin-comb}
\end{table}

\begin{table}[h]
\centering
$\begin{array}{|c|c|}
\hline
\text{SME linear combination}& \text{Value and uncertainty [GeV]}\\
 \hline
a_1   & \left(1.7 \pm  5.5\right) \times 10^{-14} \\
a_2   & \left(0.85 \pm 1.7 \right) \times 10^{-9}  \\
a_3   & \left(0.33\pm 1.2 \right)  \times 10^{-9}  \\
a_4   & \left(-1.7 \pm  7.7 \right) \times 10^{-7} \\
\hline
\end{array}$
\caption{Independent linear combinations of SME coefficients and their uncertainties (68\% confidence).}
\label{tab:lin-comb-res}
\end{table}  

We use the results of the SVD decomposition to provide an order of magnitude estimate of the so called ``maximal sensitivity'' constraints, i.e. assuming in turn that all coefficients except one are zero, and logarithmically rounding the 2$\sigma$ uncertainty \cite{kostelecky-2008}. This leads to $10^{-13}$~GeV for $\alpha(\bar{a}_{\text{eff}}^{w})_T$, $10^{-9}$~GeV for $\alpha(\bar{a}_{\text{eff}}^{w})_{Y}$, and $10^{-8}$~GeV for $\alpha(\bar{a}_{\text{eff}}^{w})_{X,Z}$, an improvement by one order of magnitude on best previous results for $\alpha(\bar{a}_{\text{eff}}^{w})_{X,Y,Z}$ \cite{flowers2017,Bourgoin2017,lshao2019_mat} and two orders of magnitude on $\alpha(\bar{a}_{\text{eff}}^{w})_{T}$ \cite{kostelecky-2008,schlamminger2008,kostelecky2011matter}.

Finally, in order to test for the possibility of hidden systematic effects we have repeated our analysis for different subsets of the data, by excluding individual measurement sessions, one at a time (so called ``jackknife'' procedure, see e.g. \cite{Bourgoin2017}). All results agree with each other and the full analysis within the uncertainties, thus providing no indication of any hidden systematics.  

\textit{Conclusion --} We have carried out a test of LLI modelled as anomalous matter-gravity couplings in the SME, and found no indication of any LLI violation. Using 5 measurement sessions of the MICROSCOPE space mission we set constraints on linear combinations of the corresponding SME coefficients $\alpha(\bar{a}_{\text{eff}}^{w})_\mu$, improving on best previous results \cite{schlamminger2008,kostelecky2011matter,kostelecky-2008,flowers2017,Bourgoin2017} by one order of magnitude on $\alpha(\bar{a}_{\text{eff}}^{w})_{X,Y,Z}$ and two orders of magnitude on $\alpha(\bar{a}_{\text{eff}}^{w})_{T}$, when assuming independence of the coefficients (so called ``maximal sensitivity'' constraints'' \cite{kostelecky-2008}). Additionally, our independent linear combinations are different from previous ones, which increases the diversity of available constraints paving the way towards a full decorrelation of the individual coefficients. More specifically, the results constrain Lorentz/WEP violation to well below to the mass scale of the electron ($10^{-3}$~GeV), thus placing severe constraints on the scenario of ``large'' Lorentz violation \cite{kostelecky2009a}. 

In the near future we expect to improve these first results, by analysing more data and by improving the estimation of the limiting systematic effect (thermal sensitivity). More data will help decorrelating the $T$ and $X$ components of $\alpha(\bar{a}_{\text{eff}}^{w})_\mu$. Indeed Fig.\ref{fig:data} shows that the sampling of the evolution of $\beta_X$ with our present data is close to a constant, hence the correlation with the $T$ component. However, the $Y$ and $Z$ components will always be strongly correlated, as the evolution of $\beta_Y$ and $\beta_Z$ is identical. Only the small terms in the second and third lines of (\ref{eq_mes_mic_sme}) will eventually allow some decorrelation. Finally, accelerations along the other two axes of the instrument were also measured and could be included in future analysis.

The MICROSCOPE mission has already provided ground-breaking results in fundamental gravitational physics, in different theoretical models \cite{Touboul2017,Berge2018,Fayet2018,Hees2018,Fayet2019}. Our first constraints in the SME presented here, add to the ever-growing literature on the results of what is one of the most successful space missions in fundamental physics so far.

\begin{acknowledgements}
The authors express their gratitude to the different partner entities involved in the mission and in particular CNES, the French space agency in charge of the satellite. This work is based on observations made with the T-SAGE instrument, installed on the CNES-ESA-ONERA-CNRS-OCA-DLR-ZARM MICROSCOPE mission. The data was provided by the MICROSCOPE ``Centre de Mission Scientifique'' at ONERA. Q.G.B. acknowledges support from NSF grant no. 1806871. G.Mo acknowledges support from the Carleton College Towsley Fund.
\end{acknowledgements}

\bibliography{biblio_microscope}
\end{document}